\documentclass[preprint,12pt]{elsarticle}

\usepackage{amssymb}

\usepackage{amsthm}

\setcitestyle{numbers, sort&compress}

\usepackage[numbers]{natbib}
\usepackage{soul}
\usepackage{booktabs} 
\usepackage{algorithm}
\usepackage[noend]{algorithmic}
\usepackage{graphicx}
\usepackage{textcomp}
\usepackage{xcolor}
\def\BibTeX{{\rm B\kern-.05em{\sc i\kern-.025em b}\kern-.08em
    T\kern-.1667em\lower.7ex\hbox{E}\kern-.125emX}}
\usepackage{hyperref}
\usepackage{amsmath,amsfonts,mathtools}

\newtheorem{definition}{Definition}[section]

\newtheorem{theorem}{Theorem}[section]

\newtheorem{corollary}{Corollary}[section]

\newtheorem{lemma}{Lemma}[section]

\usepackage{caption}
\usepackage{subcaption}

\usepackage{paralist}
\usepackage{verbatim}
\usepackage{optidef}

\definecolor{darkgrn}{rgb}{0, 0.75, 0}

\usepackage{xargs}                      
\usepackage[colorinlistoftodos,prependcaption,textsize=footnotesize]{todonotes}
\usepackage{xspace}
\usepackage[inline]{enumitem}
\usepackage{wrapfig}

\newcommandx{\ak}[2][1=]{\todo[color=green!50,#1]{\sf \textbf{AK:} #2}\xspace}

\newcommandx{\rb}[2][1=]{\todo[color=blue!10,#1]{\sf \textbf{Reet:} #2}\xspace}

\newcommandx{\mh}[2][1=]{\todo[color=green!10,#1]{\sf \textbf{MH:} #2}\xspace}

\newcommandx{\mm}[2][1=]{\todo[color=green!50,#1]{\sf \textbf{MM:} #2}\xspace}

\newcommandx{\smf}[2][1=]{\todo[color=cyan!50,#1]{\sf \textbf{smf:} #2}\xspace}

\newcommand{\greedimm}{\texttt{GreediRIS}}
\newcommand{\greedimmtrunc}{\texttt{GreediRIS-trunc}}

\newcommand{\rrrsets}{$\mathfrak{R}$}
\newcommand{\maxkcover}{\texttt{max-k-cover}}

\DeclareMathOperator*{\argmax}{arg\,max}

\def\randgreedi{\mbox{\sc RandGreedi}}
\def\greedi{\mbox{\sc Greedi}}
\def\infmax{\mbox{\sc InfMax}}

\newcommand{\cR}{\mathcal{R}}
\newcommand{\cS}{\mathcal{S}}

\usepackage[compact]{titlesec}
\titlespacing{\section}{1pt}{1ex}{1ex}
\titlespacing{\subsection}{1pt}{2ex}{1ex}
\titlespacing{\subsubsection}{1pt}{0.5ex}{0ex}

\begin{document}

\begin{frontmatter}

\title{\greedimm{}: Scalable Influence Maximization using Distributed Streaming Maximum Cover}

\author[inst1]{Reet Barik}
\author[inst1]{Wade Cappa}
\author[inst2]{S M Ferdous}
\author[inst2]{Marco Minutoli}
\author[inst2,inst1]{Mahantesh Halappanavar}
\author[inst1,inst2]{Ananth Kalyanaraman}

\affiliation[inst1]{organization={Washington State University}, 
            postcode={99164}, 
            state={WA},
            country={USA}}
\affiliation[inst2]{organization={Pacific Northwest National Laboratory}, 
            city={Richland},
            postcode={99354}, 
            state={WA},
            country={USA}}

\begin{abstract}
Influence maximization---the problem of identifying a subset of $k$ influential seeds (vertices) in a network---is a classical problem in network science with numerous applications. The problem is NP-hard, but there exist efficient polynomial time approximations. However, scaling these algorithms still remain a daunting  task due to the complexities associated with steps involving stochastic sampling and large-scale aggregations. 
In this paper, we present a new parallel distributed approximation algorithm for influence maximization with provable approximation guarantees. Our approach, which we call \greedimm, leverages the \randgreedi{} framework---a state-of-the-art approach for distributed submodular optimization---for solving a step that computes a maximum $k$ cover. \greedimm{} combines distributed and streaming models of computations, along with pruning techniques, to effectively address the communication bottlenecks of the algorithm.
Experimental results on up to 512 nodes (32K cores) of the NERSC Perlmutter supercomputer show that \greedimm{} can achieve good strong scaling performance, preserve quality, and significantly outperform the other state-of-the-art distributed implementations.
For instance, on 512 nodes, the most performant variant of \greedimm{} achieves geometric mean speedups of 28.99$\times$ and 36.35$\times$ for two different diffusion models, over a state-of-the-art parallel implementation. We also present a communication-optimized version of \greedimm{} that further improves the speedups by two orders of magnitude.
\end{abstract}

\begin{keyword}
distributed influence maximization \sep distributed submodular maximization \sep streaming maximum $k$-cover \sep parallel graph algorithms 
\end{keyword}

\end{frontmatter}

\section{Introduction}
\label{sec:intro}

Given a large real-world graph (e.g., online social network), a fixed budget $k>0$, and a stochastic diffusion model $M$, the influence maximization (henceforth, \infmax{}) problem aims to identify $k$ nodes (or ``seeds'') that when initially activated, are expected to maximize influence spread on the network under the model $M$. This ``word of mouth'' approach to spreading influence has made \infmax{} rich in applications---e.g.,  for online viral marketing \cite{domingos2001mining}, network monitoring~\cite{LeskovecKGFVG07}, controlling rumors in social networks~\cite{BudakAA11}, recommendation systems~\cite{YeLL12}, and in understanding how contagions spread in a 
population 
\cite{kempe03_maxim,azaouzi2021new}.

\infmax{} is NP-hard under classical diffusion models \cite{kempe03_maxim} such as \emph{Independent Cascade} (IC) and \emph{Linear Threshold} (LT). The seminal work by \citet{kempe03_maxim} showed that the expected influence spread function is \emph{monotone submodular}---which means the marginal gain of adding a new seed to the current solution set decreases as the set becomes larger. 
Using classical submodularity results~\cite{nemhauser1978analysis}, \citeauthor{kempe03_maxim} provided a greedy $(1-1/e)$-approximation algorithm, which incrementally expands the seed set by selecting the next seed with the highest marginal gain in influence. Although the connection to submodularity provides efficient algorithms, 
approximating \infmax{} still requires Monte Carlo simulations to generate samples that approximate the spread from the current seed set~\cite{kempe03_maxim}.

An alternative class of approaches, such as the IMM~\cite{tang2015influence} and OPIM~\cite{tang2018online} algorithms, uses the notion of \emph{reverse influence sampling} \cite{borgs2014maximizing} to obtain $(1-1/e-\varepsilon)$-approximation, where $\varepsilon\in[0,1]$ is a controllable parameter that models the sampling error. These approaches \cite{tang2014influence,tang2015influence, tang2018online} use random sampling of the graph to ``score'' vertices based on how other vertices can reach them and use that information to identify seeds.  
These algorithms and related implementations have sufficiently demonstrated that the RIS-based approach is significantly faster in practice compared to the greedy hill climbing approach \cite{kempe03_maxim}---making the RIS-based approach the {\it de facto} choice to implement \infmax.

The RIS based approaches consist of multiple rounds, each with two major phases: a) \emph{Sampling}, and b) \emph{Seed selection}. The sampling phase generates subgraph samples based on the diffusion model, and the seed selection phase solves a \emph{maximum $k$-coverage} problem on the generated samples---i.e., identifying $k$  vertex seeds that provide maximum coverage over the samples. \citet{minutoli2019fast,minutoli2020curipples} developed Ripples, the first parallel implementation for RIS-based IMM \cite{tang2015influence}.  Subsequently, \citet{tang2022distributed} developed DiIMM, which uses a similar distributed strategy to parallelize IMM. 
Both these approaches retain the approximation guarantees of their sequential predecessors.  
However, in both these methods, while the sampling step is easier to scale, the seed selection step is more challenging as it involves performing $k$ global reductions---as observed later in the results. \\

\noindent{\bf Contributions: }
 In this paper, we target the seed selection step to improve the scaling of RIS-based approaches. In particular, we propose a distributed streaming maximum-$k$-cover (abbreviated \maxkcover) to accelerate seed selection, using the distributed \greedi~\cite{MirzasoleimanBK15,BarbosaENW15} framework under a parallel setting.  In this framework, a set of local machines compute partial greedy solutions of the local dataset which they transmit to a global machine, and the global machine applies an aggregation algorithm on the partial solutions received. 

Distributed \greedi{} avoids expensive reductions. However, we observe that employing the vanilla \greedi{} to \infmax{}, the global aggregation leads to a compute and communication bottleneck as the number of machines increases. To address this issue, we have designed two techniques: a) aggregation in the global machine asynchronously by adapting a recent \emph{streaming} algorithm, and b) \emph{truncating} elements in the local partial solutions to aggressively cut down on computing and communication for the global machine. Using the streaming algorithm in the global aggregation allows us to carry out the local and global seed selections in tandem, effectively masking the communication overheads. Both of these approaches come with theoretical approximation guarantees. In summary, the main contributions are:

\vspace{-0.1cm}
\begin{itemize}[leftmargin=*,itemsep=-0.05ex]
\item 
We present \greedimm{}, a new distributed streaming algorithm for \emph{RIS}-based \infmax{} on distributed parallel platforms (\S\ref{sec:GreeDIMM}). 
\item We present and implement \greedimmtrunc{}, a truncation technique, to provide a knob to control communication overhead with a tradeoff in quality (\S\ref{sec:truncated}).
\item All our algorithms have provable worst-case approximation guarantees (\S\ref{sec:truncated}).
\item Experimental evaluation on up to 512 compute nodes (32K cores) of a supercomputer demonstrates that  \greedimm{} significantly outperforms the state-of-the-art distributed methods~\cite{minutoli2019fast,tang2022distributed}  (reported speedups with geometric mean of 28.99$\times$ and 36.35$\times$ for two popular diffusion models), while preserving output quality (geometric mean of reported quality change from state-of-the-art baseline is $2.72\%$). 
\end{itemize}
 
Our work---the first of its kind to incorporate distributed parallel submodular optimization for \infmax---is also generic enough to be extended to a broader base of application settings that use submodular optimization. While such adaptations need significant efforts in implementations, this paper represents a concrete demonstration for \infmax.

\section{Background and Preliminaries}
\label{sec:background}

In this section, we review the necessary background for the \infmax{} problem, including submodular optimization, \maxkcover, and the IMM algorithm which forms the sequential template for our method. 
Table~\ref{tab:notation} lists the key notations used in the paper.

\begin{table}[tbh]
\small
\centering
    \vspace{-0.2cm}
    \caption{Notation}
    
    \centering
    \begin{tabular}{l|l}
         \toprule
         \textbf{Symbol} & \textbf{Description} \\
         \midrule
         $G=(V, E)$ & Input graph with $n$ vertices in $V$ \\
         $S$ & Seed set, $S \subseteq V$\\
         $k$ & Target number of output seed vertices  \\
         $M$ & Diffusion process for influence spread over $G$\\
         $\sigma(A)$ & Expected influence of $A \subseteq V$ under $M$\\
         $OPT$ & Max influence spread of any set of $k$ vertices\\
         $N_a(u)$ & Subset of activated neighbors of vertex $u$ \\
         \hline
         $RRR(u)$ & Random Reverse Reachable (RRR) set for $u$\\
         $\theta$ & Target number of RRR sets (i.e., samples)\\
         \rrrsets{} & Set containing $\theta$ RRR samples\\
         $\cR(i)$ & Subset of vertices in $V$ present in RRR set $i$\\
         $\cS(v)$ & Covering subset for $v$: $\cS(v)=\{i|v\in\cR(i)\}$\\
         $m$ & Number of machines (or processes)\\
         \bottomrule
    \end{tabular}
    \label{tab:notation}
\end{table}

Let  $G=(V,E)$ be a graph, where $V$ and $E$ are the sets of vertices and edges, respectively, and $n=|V|$. 
The process of influence spread on $G$ can be described as follows. 
Let $S\subseteq V$ denote a \emph{seed set} of vertices which are already \emph{activated}. 
Then, the \emph{expected influence of $S$} (denoted by $\sigma(S)$) is the expected number of vertices that will be activated by $S$ through a stochastic diffusion process, defined by a model $M$. 
Note that $0<\sigma(S)\leq n$.
\begin{definition}[\infmax]
Given a graph $G=(V,E)$, a diffusion process $M$, and an integer $k$, the Influence Maximization problem finds a seed set $S\subseteq V$, where $|S|\leq k$ maximizing $\sigma(S)$ under $M$.
\end{definition}

\citet{kempe03_maxim} showed that \infmax{} is NP-hard under two practical diffusion models, namely, Independent Cascade (IC) and Linear Threshold (LT).
In the IC model, at each diffusion step, every edge $\langle u, v\rangle$ has a fixed probability $p_u(v)$ for an active vertex $u$ to activate a neighbor $v$.
The process stops when no further activation is possible. 
The LT model describes group behavior where individuals (i.e., $v \in V$) have some inertia (modeled by a scalar $\tau_v$) in adopting mass behavior.  The strength of the relationship between two vertices $u, v \in V$ is captured by a probabilistic edge weight $w_{(u,v)}$, and the sum of the incoming weights on any vertex is assumed to be 1.
A vertex $v$ becomes active in LT  if $\sum_{u \in N_a(v)} w_{u,v} \geq \tau_v$, where $N_a(v)$ is the subset of active neighbors of $v$.
\citet{kempe03_maxim} also proved that the expected influence $\sigma(S)$ is a non-negative monotone submodular function. Submodular functions model  diminishing returns of utilities. More formally:
\begin{definition}[Submodular Function]
Let $X$ be a finite set, the function $f: 2^X \to \mathbb{R}$ is submodular if and only if:
\begin{equation}
    f(A \cup \{x\}) - f(A) \geq f(B \cup \{x\}) - f(B),
\end{equation}
where $A \subseteq B \subseteq X$ and $x \in X \setminus B$.
\end{definition}
For \infmax, where $X=V$, the left term represents the \emph{marginal gain} in influence for adding a new vertex $x$ to an existing seed set $A\subseteq V$.
By the submodularity property, the net gain of adding a new vertex into the current seed set could only diminish as the seed set expands---suggesting a greedy approach toward seed set expansion. 
This seminal result led to the first greedy approximation algorithm for \infmax{}~\cite{kempe03_maxim}---using greedy hill climbing~\cite{nemhauser1978analysis}. 
Estimating the actual influence spread  
is also shown to be \#P-hard under both the IC and LT models~\cite{ChenWW10,ChenYZ10}. The existing algorithms for \infmax{} thus employ randomized sampling to estimate expected influence and introduce an additive sampling error in the approximation factor. The early efforts focused on simulation-based approaches such as Monte-Carlo simulation~\cite{kempe03_maxim}.
Subsequently, another class of approximation algorithms (albeit faster) was developed by exploiting \emph{Reverse Influence Sampling (RIS)} that uses the notion of reverse reachability \cite{borgs2014maximizing}. 
More specifically, this strategy is based on the construction of \emph{random reverse reachable} set for vertices.
\begin{definition}[Random Reverse Reachable (RRR) set]
Let $g$ be a random subgraph of $G$, obtained by removing edges randomly as dictated by their respective edge probabilities.
The \emph{random reverse reachable} set for vertex $u\in V$ in $g$ is given by: 
$RRR_g(u) = \{ v | \exists \textrm{ a path from } v\textrm{ to }u\textrm{ in }g.\}$
\end{definition} 
We refer to each $RRR_g(u)$ as a \emph{sample} generated for $u$.
Note that each such sample is a subset of $V$. 
An alternative interpretation of $RRR_g(u)$ is---if a vertex $v$ appears in $RRR_g(u)$, then $v$ is a likely influencer of $u$. Consequently, higher the frequency of the vertex $v$ appearing in such random reverse reachable sets, the larger is its expected influence in the input network.

\subsection{The IMM algorithm} 
\label{sec:IMM-background}

\citet{tang2015influence} proposed the IMM algorithm to implement the RIS approach~\cite{borgs2014maximizing,tang2014influence}.
The algorithm is summarized in Algorithm~\ref{alg:imm-serial}. 
The main intuition is to generate $\theta$ number of RRR samples, and then select a set of $k$ vertices as the solution seed set such that they provide maximum coverage over the set of samples. It uses Martingale analysis and bootstrap techniques to estimate a lower bound of the sampling effort required to achieve OPT (the utility of the intractable optimal solution). 
Initially, using the analytical formula prescribed in \cite{tang2015influence}, the algorithm estimates the number of RRR samples to compute, denoted by $\hat{\theta}$, using the values of $|V|$, $\varepsilon$ and $k$ (\texttt{Estimate(.)} in line 3).
Next, the sampling procedure \texttt{Sample(.)} generates and populates those $\hat{\theta}$ RRR samples in \rrrsets{} (line 7). To generate such a sample:
\begin{enumerate}[leftmargin=*,noitemsep,topsep=0pt,parsep=0pt,partopsep=0pt]
\item a vertex $u \in V$ is chosen uniformly at random to be the source vertex.
\item given a source, a probabilistic BFS (edges are selected for traversal based on their assigned weights) on the reverse graph (destinations are vertices with incoming edges to the source) is used to populate the RRR sample.
\end{enumerate}
The generated samples are then used to compute $k$ seeds by the seed selection step \texttt{SelectSeeds(.)} as shown in the line 8. This is an instance of computing a \maxkcover{} on the universe of samples. Using this intermediate seed set, (provably an unbiased estimator of the unknown optimum (OPT)) a lower bound (LB) on OPT is computed, and compared against a certain fraction of $|V|$ (denoted by function \texttt{CheckGoodness} in Algorithm~\ref{alg:imm-serial}; details in  \cite{tang2015influence}). If the condition isn't met, $\hat{\theta}$ is doubled and the process is repeated. The IMM algorithm conducts at most $\log{|V|}$ such rounds, which we will term as ``martingale'' rounds (lines 4-11) till the condition is met. Once a tight lower bound is computed, the final  sample size ($\theta$) is calculated using the LB, and the martingale loop is terminated. 
Finally, $\theta$ RRR samples are generated, and the set of $k$ seeds are selected using those samples (lines 12-13).

The IMM algorithm is shown to be $(1-1/e-\varepsilon)$-approximate for \infmax. Here, $\varepsilon$ is the precision parameter dictating the sampling error. We restate a result from \cite{tang2015influence} here.
\begin{theorem}
\label{thm:imm-max-cover}
Given $\theta = \lambda^*/OPT$ and $\delta \in (0,1)$, the Algorithm~\ref{alg:imm-serial} returns an $(1-1/e-\varepsilon)-$approximate solution of \infmax\ with probability at least $1-\delta$.
\end{theorem}

Here $OPT$ is the optimal expected influence. The $\lambda^*$ is a function of $n,\varepsilon,k, \delta$. We refer to~\cite{tang2015influence,chen2018issue} for the actual value of the function $\lambda^*$. The $(1-1/e)$ term in Theorem~\ref{thm:imm-max-cover} is due to the standard greedy algorithm for the seed selection that computes a \maxkcover{} on the universe of $\theta$ samples. In fact, we can replace the greedy algorithm with any $\alpha$-approximate algorithm, where $\alpha \leq (1-1/e)$.

\begin{algorithm}[tbh]
\caption{IMM($G,k,\varepsilon$) by \cite{tang2015influence}}
\label{alg:imm-serial}
\begin{algorithmic}[1]
\REQUIRE{$G(V,E)$: graph; $k$: \# of seeds; $\varepsilon\in[0,1]$: sampling error.}
\ENSURE{$S$: the seed set}

\STATE 
\COMMENT{\textcolor{blue}{1. Estimate $\theta$  using martingale rounds}}
\STATE \rrrsets $\gets \emptyset$
\STATE $\hat{\theta}\gets$ \texttt{Estimate}$(k,\varepsilon,|V|)$
\FOR{$x\in [1,\log |V|]$}
    \IF{$x > 1$}
        \STATE $\hat{\theta}\gets 2\times\hat{\theta}$
    \ENDIF
    \STATE \rrrsets $\gets$ \texttt{Sample}$(G,\hat{\theta}-|$\rrrsets$|,$\rrrsets$)$
    \STATE $S\gets$ \texttt{SelectSeeds}$(G,k,$\rrrsets$)$ \label{ln:seed-select1}
    \IF{\texttt{CheckGoodness($S,V,\& LB$)}}
        \STATE $\theta\gets f(k,\varepsilon,|V|,LB)$ 
        \STATE \textbf{break} \COMMENT{\textcolor{blue}{lower bound condition is met}}
    \ENDIF
\ENDFOR

\STATE \rrrsets$\gets$ \texttt{Sample}$(G,\theta)$ \COMMENT{\textcolor{blue}{2. Generate $\theta$ RRR samples}}
\STATE $S\gets$ \texttt{SelectSeeds}$(G,k,$\rrrsets$)$ \COMMENT{\textcolor{blue}{3. Select final seeds}}\label{ln:seed-select2}
\RETURN S
\end{algorithmic}
\end{algorithm}

\begin{corollary}
\label{cor:imm-max-cover}
Given $\theta = \lambda^*/OPT$, $\delta \in (0,1)$ and $\alpha$-approximate \maxkcover, the IMM algorithm returns an $(\alpha-\varepsilon)-$approximate solution of  \infmax\ with probability at least $1-\delta$.
\end{corollary}

The justification of the Corollary~\ref{cor:imm-max-cover} is as follows. Since the sampling effort, $\theta = \lambda^*/OPT$ is estimated to guarantee an $(1-1/e)$-approximation for the $\maxkcover$, for any $\alpha \leq (1-1/e) $, the sampling effort is trivially sufficient.

The \texttt{Sample(.)} function is usually implemented as a probabilistic BFS originating from $\hat{\theta}$ randomly selected roots from $V$ in any martingale round. 
 
All vertices visited in a probabilistic BFS become the RRR set for that sample. 
The \texttt{SelectSeeds(.)} procedure is greedy, and involves iteratively computing $k$ vertices with a maximum coverage on the 
set of RRR samples as detailed in \S\ref{sec:distIMM-base}.

Among distributed parallel approaches for \infmax, the state-of-the-art comprises of PREEMPT \cite{minutoli2020preempt} and IMpart \cite{barik2022impart} for greedy hill climbing, and Ripples \cite{minutoli2019fast,minutoli2020curipples} and DiIMM \cite{tang2022distributed} for parallel distributed IMM.
As we focus on RIS-based approaches, we will review Ripples and DiIMM. \\

\noindent{\bf Prior work in parallel distributed IMM: }
Both Ripples and DiIMM approaches start by loading the entire graph on each machine, and efficiently distributing the generation of all ($\theta$ or $\hat{\theta}$) samples across $m$ machines, i.e., $O(\frac{\theta}{m})$ samples per machine.

However, seed selection involves significant data exchange to compute the updated marginal coverage for each vertex, making the step communication-intensive.  In particular, since sampling is distributed, each machine can keep track of only the local sample coverage for each of the $n$ vertices. Thus,to get the global coverage values, a global reduction is required each time a new seed is selected. The Ripples algorithm implements this using $k$ global reductions (over an $n$-sized frequency vector).
The updated global frequencies are used to pick the next best seed.
In contrast, DiIMM  computes these updates in a lazy fashion under a master-worker setting. Here, after performing a global reduction to select the first seed, the master processes the remaining vertices in a non-increasing order of their original coverage. 
If the next vertex selected is detected to have an outdated coverage, it is pushed back into the queue and the algorithm continues. Once it finds the next seed (one with the correct next best marginal coverage), it broadcast the information to all the machines so that they can update vertex coverages using their local sample sets. The master machine accumulates the changes via a global reduction for the next iteration of seed selection.

We can show that the DiIMM approach is algorithmically equivalent of performing $k$ global reductions. While Ripples does this entirely in a distributed fashion, DiIMM computes this using a master-worker setting.
In either case, seed selection dominates communication costs, as it entails a reduction of an $O(n)$-sized vector within each of the $k$ iterations.

In our work, the distributed sampling step is similar to Ripples and DiIMM---as it is a step that is easily amenable for distributed parallelism.
The major difference, however, stems in the way we perform seed selection. More specifically, we reformulate the seed selection step using the \randgreedi{} framework \cite{BarbosaENW15} so as to bring the state-of-the-art for distributed submodular optimization to parallelize influence maximization. 
The key advantages are: a) the $k$-steps of communication can be reduced to just two steps (as will be explained next in \S\ref{sec:distIMM-base}); and b) streaming also can be used (along with a truncation scheme) to mask out communication overheads further (as will be explained in \S\ref{sec:distIMM-streaming}).

\section{\greedimm: A \randgreedi\ Framework for Distributed Influence Maximization}
\label{sec:GreeDIMM}

We exploit the submodular nature of the \maxkcover{} computation step in order to accelerate seed selection for \infmax.  
The state-of-the-art distributed submodular maximization algorithm is the \randgreedi{}~\cite{BarbosaENW15} framework. 
Adapting this framework, we present a new distributed memory parallel algorithm for influence maximization and refer to our approach \greedimm. 

We first provide a brief overview on \greedi{} and \randgreedi{} (\S\ref{sec:greedi-background}). Then we show how computing \maxkcover{} allows adapting \randgreedi{} for any \emph{RIS}-based \infmax{} algorithms(\S\ref{sec:distIMM-base}). We then discuss the limitations of this simple adaptation and propose a distributed streaming algorithm (\S\ref{sec:distIMM-streaming}) and a communication-reducing truncation technique (\S\ref{sec:truncated}) to alleviate these limitations.
Section~\ref{sec:parallel} describes our overall parallel algorithm.

\subsection{The \greedi{} framework}
\label{sec:greedi-background}
The \greedi{} framework represents approaches \cite{BarbosaENW15, MirzasoleimanKSK13,MirrokniZ15} for distributed submodular function optimization under budget constraints. 
These approaches initially partition the data set into different machines. Each machine, independently in parallel, executes the standard greedy algorithm~\cite{nemhauser1978analysis} to compute its local solutions. These local solutions are then communicated to a global aggregation node, where they are merged. The greedy algorithm is then employed in the global machine on this set of merged local solutions. The final solution is the one that maximizes the function value from the set of local solutions and the global solution. This \emph{two-stage} algorithm is shown to be more efficient~\cite{BarbosaENW15} than earlier distributed algorithms.  
Although  \greedi{} performs  well in practice, its approximation ratio depends inversely on the number of machines. \randgreedi{} \cite{BarbosaENW15} was proposed to improve the approximation guarantee of \greedi. Through a probabilistic analysis, the authors showed that by partitioning the data uniformly at random,  \greedi{} could achieve a constant approximation guarantee. 
We restate the following theorem from~\cite{BarbosaENW15} that proves the approximation ratio of \randgreedi.

\begin{theorem}
\label{thm:randgreedi}
Let the greedy method on the local machines be an $\alpha$-approximate, and that the global algorithm be a $\beta$-approximate for maximizing a monotone submodular function subject to a cardinality constraint.
Then, \randgreedi{} is (in expectation) $\frac{\alpha \beta}{\alpha+\beta}$-approximate for the same problem.
\end{theorem}

While the local machines in \randgreedi{} need to execute the greedy algorithm locally, the global algorithm can use any approach.

\subsection{Distributed IMM using \randgreedi}
\label{sec:distIMM-base}

We first extend IMM  into the \randgreedi{} framework.
The focus on IMM is for expository reasons, as it a prototypical RIS approach. Our distributed approach also works for OPIM \cite{tang2018online}.

Recall from Table~\ref{tab:notation} that \rrrsets{} denotes the collection of sampled $\theta$ RRR sets, and $\cR(i) \subseteq V$ denote the $i$th RRR set in this collection. 
Given \rrrsets, $V$, and the target number $k$, the seed selection problem can be formulated as a \maxkcover{} computation as follows.
The set \rrrsets{} represents the \emph{universe} to be covered. 
From \rrrsets, we generate a collection of \emph{covering subsets}, $\cS$. 
For each node $v \in V$, let $\cS(v)$ be the subset of RRR sets (identified by their respective indices in $[0,\theta-1]$) in which $v$ appears---i.e.,  $\cS(v) = \{i \in[0,\theta-1]| v \in \cR(i)\}$. It is now easy to see that the seed selection problem of IMM reduces to finding a \maxkcover, where the universe is $\{0,\ldots,\theta-1\}$ (i.e., \rrrsets), and the collection of covering subsets is $\cS$. More formally, the \maxkcover{} problem here is to find a subset $S\subseteq V$  that is given by: 
$\textrm{arg\ }\max_{S \subseteq V} C(S) = \left |\bigcup_{i \in S} \cS(i) \right |\textrm{, subject to }|S| \leq k.$

The function $C(S)$ can easily be shown to be non-negative monotone submodular. A greedy algorithm that repeatedly selects a covering subset from $\cS$ with the largest marginal gain with respect to the current solution, achieves $(1-1/e)$  worst-case approximation guarantee~\cite{nemhauser1978analysis} which is the best achievable unless P=NP~\cite{Feige98}. 
In practice, faster variants of greedy exist, such as lazy greedy~\cite{minoux1978accelerated}, threshold greedy~\cite{GuptaRST10,BadanidiyuruV14}, and stochastic greedy~\cite{MirzasoleimanBK15}.
Lazy greedy~\cite{minoux1978accelerated}, which exploits the monotone decrease in the marginal gain of a submodular function, is widely used and often performs much faster than the standard greedy in practice \cite{BadanidiyuruV14} while obtaining the same approximation guarantee. 
Our implementation supports both standard and lazy greedy approaches. Algorithm~\ref{lz-max-k-cover} shows our lazy greedy \maxkcover.

\begin{algorithm}
\caption{Lazy-greedy-max-k-cover($\hat\theta,\cS,k$)}
\label{lz-max-k-cover}
\begin{algorithmic}[1]
\REQUIRE{$\hat\theta$: size of the universe, $\cS$: $n$ covering subsets, $k$: an integer}
\ENSURE{$S$: A subset of $[0,\ldots,n-1]$ of size at most $k$}
\STATE Build a max heap $Q$ using all $n$ covering subsets $\cS$, and using the cardinality of each subset as the key

\STATE $S = \emptyset$

\WHILE{$(|S| < k)$ or Q is empty}
\STATE $v$ = Q.pop()
\STATE marginal\_gain = $C(S \cup \{v\}) - C(S)$

\IF{marginal\_gain $\geq$ Q.top().key}
\STATE $S = S \cup \{v\}$
\ELSE
    \STATE Q.push($\langle$$v$,marginal\_gain$\rangle$)
\ENDIF
\ENDWHILE
\RETURN $S$
\end{algorithmic}
\end{algorithm}

Next, we describe our distributed parallel version of Algorithm~\ref{alg:imm-serial} using the \randgreedi\ framework. 
Let $m$ to denote the number of machines (or processes or ranks, used interchangeably), and $p\in[0,m-1]$  denote an arbitrary machine.
Given $G(V,E)$, $k$, and $m$, our distributed IMM algorithm conducts multiple martingale rounds just as the sequential version (Algorithm~\ref{alg:imm-serial}). 
Algorithm~\ref{alg:Randgreedi-IMM} shows the \randgreedi{} version of our distributed IMM  for each round.

During the \textbf{sampling} phase (lines 1-6 of Algorithm~\ref{alg:Randgreedi-IMM}), each machine independently generates $\hat\theta/m$ samples from $G$. This is achieved by selecting vertices in $V$ uniformly at random and generating $RRR(v)$ for each of them. 
We use the Leap Frog method described in \cite{minutoli2019fast} to ensure consistent parallel pseudorandom generation across different values of $m$. 
At rank $p$, the parallel sampling phase produces $\mathfrak{R}_p$.
Since $\hat\theta$ doubles with each martingale round, in our implementation we retain the previous batch of samples and simply add the second half (i.e., augment the previous round's set of $\frac{\hat\theta/m}{2}$ samples with a new set of $\frac{\hat\theta/m}{2}$ samples).
The samples in $\mathfrak{R}_p$ are numbered from $[p\cdot \frac{\hat\theta}{m}, (p+1) \cdot \frac{\hat\theta}{m}-1]$, so that each rank can claim a disjoint interval.
Next, we generate a partitioning of the set of vertex ids $[0,n-1]$ uniformly at random across the $m$ machines so that each rank $p$ is assigned a distinct subset of $\approx n/m$ ids (line 7 of Algorithm~\ref{alg:Randgreedi-IMM}). Let $V_p\subseteq [0,n-1]$ denote this subset assigned to rank $p$. 
Note that as this is a uniform random partitioning, and therefore each vertex is assigned only to one rank.

\begin{algorithm}[tbh]
\caption{\label{alg:Randgreedi-IMM}
\randgreedi-RIS-Round($G,k,\hat\theta,m$)}
\begin{algorithmic}[1]
\REQUIRE{$G(V,E)$: input graph, $k$: an integer, $\hat\theta$: number of samples for the current round, $m$: number of machines}
\ENSURE{$S$: the seed set}

\FOR{ each machine $p \in [m]$ {\bf in parallel}}

\STATE $\mathfrak{R}_p\gets$ Generate $\hat\theta/m$ random samples  
    \FOR{each vertex $v \in V$}
    \STATE $\cS_p(v) = \{j | v \in \cR_p(j)\}$ 
    \ENDFOR 
\ENDFOR

\STATE Partition $V$ uniformly at random into $m$ vertex sets ($V_p$ at  $p$)

\STATE All-to-all: $\cS(u) = \bigcup_{p \in [m]}\{\cS_p(u) | u \in V_p\}$ at $p$

\STATE $S \gets$ \randgreedi-max-k-cover($\hat\theta,\cS,k,\{V_p\}$)
\RETURN $S$
\end{algorithmic}
\end{algorithm}

\begin{wrapfigure}{l}
{0.50\textwidth}

    \centering
   \includegraphics[scale=0.30]{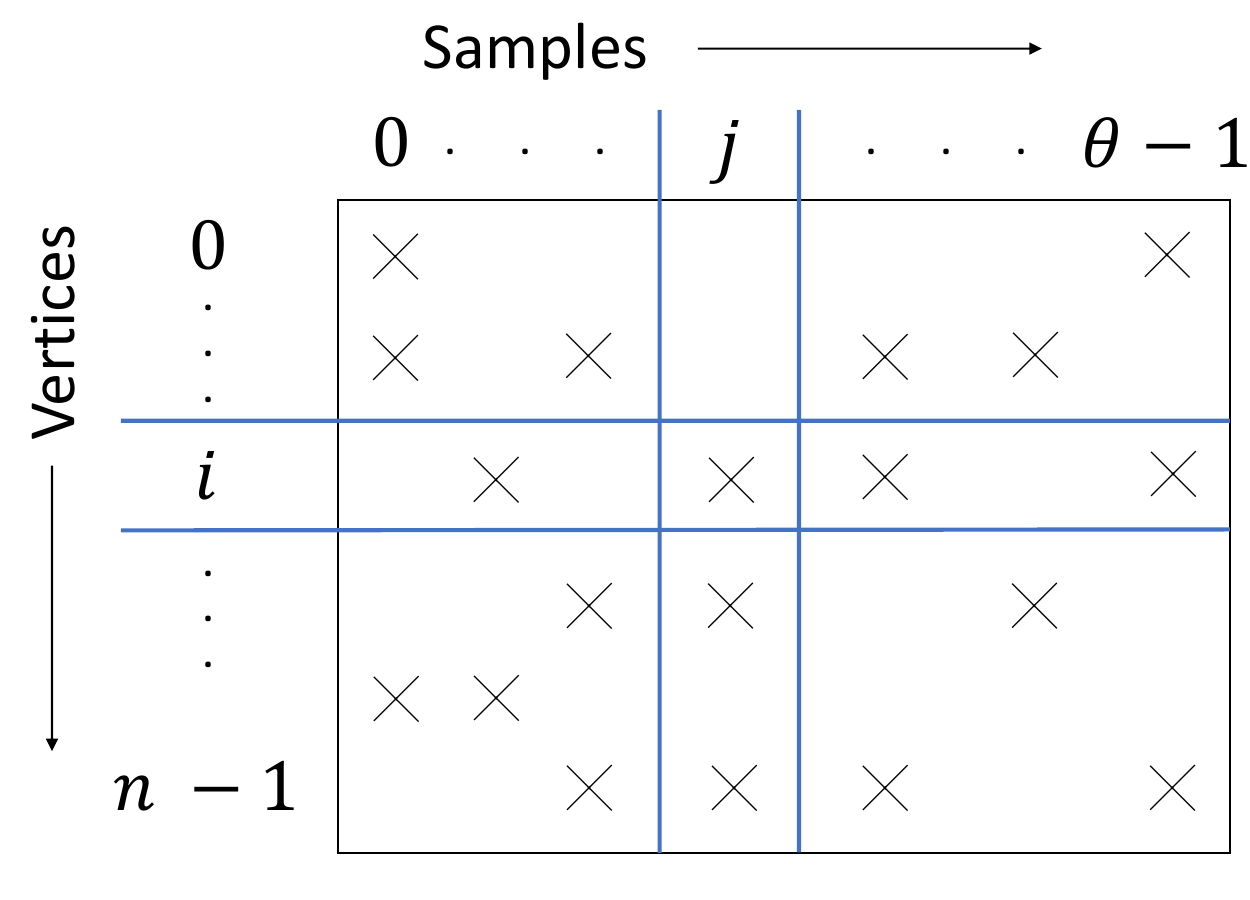}

    \caption{\small Sampling visualized as a sparse matrix. 
    The sampling phase populates the columns, and the shuffle phase distributes by rows. 
    }
    \label{fig:All2AllSchematic}
\end{wrapfigure}

Next, we perform a \emph{shuffle} communication such that the \emph{ids} of all the RRR sets that contain a particular vertex $u$ are gathered at the rank that is responsible for $u$. This  is implemented using an MPI \texttt{all-to-allv} communication primitive (line 8 of Algorithm~\ref{alg:Randgreedi-IMM}). Specifically, we first construct a set of partial covering subsets $\cS_p(u)$ at each rank $p$ using the RRR sets of $\mathfrak{R}_p$ (line 4 of Algorithm~\ref{alg:Randgreedi-IMM}). These partial covering subsets are then scattered so that the lists corresponding to all vertices $u\in V_p$ are gathered at rank $p$ as follows:
$\cS(u) = \bigcup_{p \in [m]}\{\cS_p(u) | u \in V_p\}$ at $p$.
At the end of this step, rank $p$ contains the \textit{complete} covering subset $\cS(u)$ for each $u\in V_p$.
Figure~\ref{fig:All2AllSchematic} illustrates this shuffle operation.

We are now ready to apply the \randgreedi{} framework to \textbf{compute seeds}. Algorithm~\ref{alg:Randgreedi-maxkcover} shows the function to compute the seed set using \maxkcover. As explained in \S\ref{sec:greedi-background}, the \randgreedi{} framework involves two steps: computing local \maxkcover{} in a distributed fashion, and then merging those local solutions into a global (single) machine to compute a global \maxkcover. Subsequently, the global solution ($S_g$) is compared with the best of the local solutions ($S_\ell$), and the best of the two is output as the final solution ($S$) for that martingale round. This is followed by a broadcast call to make the utility of the selected seed set available at all machines.

\begin{algorithm}
\caption{\label{alg:Randgreedi-maxkcover}
\randgreedi-max-k-cover($\theta,\cS,k,\{V_p\})$}
\begin{algorithmic}[1]
\REQUIRE{$\hat\theta$: size of the universe, $\cS$: $n$ covering subsets, $k$: an integer, $\{V_p\}$: uniform random set of partitions of $V$}
\ENSURE{$S$: A subset of $V: [0,\ldots,n-1]$ of size at most $k$}

\FOR{each process $p \in [m]$ {\bf in parallel}}

\STATE $S_p\gets$ Greedy-max-k-cover($\hat\theta,\cS_p,k$) \textcolor{blue}{//local \maxkcover}
\ENDFOR
\STATE Gather: $\cS^\prime = \bigcup_p \cS(S_p)$
\STATE $S_g\gets$ Lazy-greedy-max-k-cover($\hat\theta,\cS^\prime,k$) \textcolor{blue}{//global \maxkcover}
\STATE $S_\ell = \argmax_{p \in [m]}\{C(S_p)\}$ \textcolor{blue}{//local best}
\STATE $S = \argmax\{C(S_g), C(S_\ell) \}$
\RETURN $S$
\end{algorithmic}
\end{algorithm}

\noindent{\bf Limitations:}
A straightforward implementation of Algorithms~\ref{alg:Randgreedi-IMM} and \ref{alg:Randgreedi-maxkcover} could lead to
the global aggregation step becoming a bottleneck, as it has to wait until all the local machines complete their local \maxkcover{} computations. 

\begin{table}[h]

     \caption{\small The runtimes in seconds for computing the local and global solutions of \maxkcover{} using the \randgreedi{} template. Livejournal ($|V| = 4.8M, |E| = 68.9M$) was used as the test input.}
     \vspace{-0.0cm}
    \centering
    \begin{tabular}{l|r|r|r|r|r|}
     & \multicolumn{5}{|c|}{$m$: Number of nodes} \\ \hline
    Time (in sec) & \textit{8} & \textit{16} & \textit{32} & \textit{64} & \textit{128} \\ \hline
    local \maxkcover{} & 1.87 & 0.91 & 0.34 & 0.17 & 0.10 \\
    global \maxkcover{} & 0.22 & 0.67 & 1.20 & 2.47 & 4.86 \\ \hline
    \end{tabular}
 
    \label{tab:Naive:prelimresult}
\end{table}

With increasing $m$, this issue is exacerbated  by the global aggregation step receiving up to $m\cdot k$ local solutions.  
To test this hypothesis, we implemented and tested this template version.
Table~\ref{tab:Naive:prelimresult} shows the running times for the local and global \maxkcover{} steps as a function of $m$. The results show that as $m$ is increased, the time to generate the local solutions decreases while the time to aggregate and produce the global solution increases. 
Also, the global aggregation step chooses the final $k$ seeds from a significantly richer set of candidates, thereby increasing the amount of computations required to calculate the marginal gains to select each seed. 
This motivates the design of a streaming based implementation presented next in \S\ref{sec:distIMM-streaming}.

\subsection{Distributed Streaming IMM via \randgreedi}
\label{sec:distIMM-streaming}

To overcome the above limitations of \randgreedi, we present two improvements. 
The first approach (\S\ref{sec:streaming}) replaces the global aggregation algorithm with a streaming \maxkcover{} technique inside each round. 
The second approach (\S\ref{sec:truncated}) reduces communication cost from the local \maxkcover{} solvers through a truncation technique.

\subsubsection{Streaming computation in the global machine}
\label{sec:streaming}
The first approach to improve \randgreedi{} replaces the global aggregation algorithm with a \emph{streaming} \maxkcover{}. 
The key idea is to immediately send the local seeds to the global machine, as they are generated, without waiting for the local machines to complete all their greedy selections. The global machine then runs a linear time one-pass streaming algorithm on the incoming data, thus offsetting the computational expense, and allowing masking of communication overhead.
Streaming can therefore allow tandem executions of global and local solvers.explained in \S\ref{sec:parallel}). 

Implementing streaming necessitates that the global \maxkcover{} algorithm be changed to work with the incoming stream of local seeds (from different machines). The standard greedy \maxkcover{} algorithm (Algorithm~\ref{lz-max-k-cover}) being offline, is not suitable here.  However, with the \randgreedi\ framework, we are free to choose any algorithm for aggregation.
As the \maxkcover{} problem is an instance of the submodular maximization with cardinality constraints, 
we can use one of the recent direct streaming algorithms designed for \maxkcover{} \cite{McGregorV19,BateniEM17}.  

In this work, we employ the $(1/2-\delta)$-approximate streaming algorithm (described next) also developed in~\cite{McGregorV19} because it's nearly linear in runtime, and absence of any post-processing steps, thus generating the solution immediately after the streaming phase ends. 
The challenge here is also to show that the streaming algorithm provides good quality solution efficiently.

Let $u$ and $l$ be an upper and lower bound on optimum coverage, respectively. For a given $0<\delta < 1/2$, we maintain $B = \log_{1+\delta}\lfloor\frac{u}{l}\rfloor $ buckets indexed by the integers $[0,B-1]$. Each bucket $b$ has a lower value corresponding to the lower threshold limit representing the bucket, which is $(1+\delta)^b$.  

When a streamed-in seed along with its covering subset $s$ arrives, the algorithm computes the marginal gain of $s$ with respect to current solution of $b$, $S_b$. If the bucket $b$ holds a solution set of size smaller than $k$ and the marginal gain is greater than $(1+\delta)^b$, the vertex representing the subset$s$ (denoted by id(s)) is inserted into the current solution, and the cover set ($C_b$) is updated; otherwise, $s$ is discarded. This operation is repeated for all the buckets. 
Since the decision to recruit a seed into a bucket is independent across buckets, we use multithreaded parallelism to process the buckets. 
Once the streaming phase ends, we output the solution in the bucket ($b^*$) with maximum coverage. Algorithm~\ref{alg:stream-cover} runs in $O(\theta B|\cS_m|/t)$ time (where $t$ is the number of threads), and is shown to be $(1/2-\delta)$-approximate of the optimal coverage~\cite{McGregorV19}. 

\begin{algorithm}
\caption{Streaming \maxkcover{} at the global receiver
}
\label{alg:stream-cover}
\begin{algorithmic}[1]
\REQUIRE{$\cS_m$: streaming collection of subsets from local senders, $k$: an integer, $u$: upper bound on OPT, $l$: lower bound on OPT, $\delta$}
\ENSURE{set of seeds of size at most $k$}

\STATE{Create $B = \log_{1+\delta}\lfloor\frac{u}{l}\rfloor $ buckets}
\STATE{$C_b \gets $ covering sets of a bucket $b$, initialized to empty}
\STATE{$S_b \gets $ set of seed nodes at bucket $b$, initialized to empty}

\FOR{$s \in \cS_m$}
\FORALL{$b\in [0,B-1]$ {\bf in parallel}}
\IF{$ |S_b| < k$ and $|s \setminus C_b| \geq \frac{(1+\delta)^b}{2k}$}
    \STATE{$C_b = C_b \cup \{s\}$}
    \STATE{$S_b = S_b \cup id(s)$}
\ENDIF
\ENDFOR
\ENDFOR

\STATE{$b^* = arg \max_{b}{|C_b|}$}

\RETURN{$S_{b^*}$}
\end{algorithmic}
\end{algorithm}

\begin{lemma}
\label{lem: grdimm-strm}
The approximation ratio of  \greedimm\ with streaming aggregation is $\frac{(1-1/e) (1/2-\delta)}{(1-1/e) +(1/2-\delta)} -\varepsilon$ in expectation.
\end{lemma}
\begin{proof}
Since the local machines run a greedy algorithm the approximation guarantee is $(1-1/e)$. The streaming algorithm is $(1/2-\delta)$-approximate. The guarantee follows from Theorem~\ref{thm:randgreedi} and Corollary~\ref{cor:imm-max-cover}.
\end{proof}

\subsubsection{Communication reduction using truncation at the senders}\hfill
\label{sec:truncated}

In the \randgreedi{} framework for IMM, each machine can send up to $k$ seeds, implying a total of $m\cdot k$ seeds arriving at the global machine (be it with or without streaming). 
Our second improvement to \randgreedi{} aims at reducing the volume of communication by restricting the local senders to send only a subset of the $k$ seeds to the global machine. 
We refer to this approach as \textit{truncated greedy}. 
Specifically, each local machine (sender) still computes all $k$ local seeds. However, during streaming each sends only the top $\alpha\cdot k$ seeds (along with their local covering sets), where $0 < \alpha \leq 1$. 
We show that the approximation ratio for the truncated greedy is $(1-e^{-\alpha})$ compared to the optimal solution of $k$ subsets. Thus, the $\alpha$ value provides a trade-off between the scalability and approximation of the overall \infmax\ algorithm. 

\begin{lemma}
\label{lem:res-grdy}
The approximation guarantee of truncated greedy is $1-e^{-\alpha}$.
\end{lemma}

\begin{proof}
    We modify the standard analysis of the greedy algorithm for \maxkcover{} ~\cite{Feige98}. Let $OPT$ be the optimal coverage of the $k$ elements, and $x_i$ be the number of new elements covered by the truncated greedy algorithm in the $i$-th set it selects. We are interested to bound the coverage value of the truncated greedy, i.e., $\sum_{j=1}^{\alpha k} x_j$. Since the optimal algorithm uses $k$ sets to cover $OPT$ elements, at iteration $(i+1)$ of the algorithm, there must be a set that covers at least $\frac{OPT - \sum_{j=1}^i x_i}{k}$ new elements. So, $x_{i+1} \geq \frac{OPT - \sum_{j=1}^i x_i}{k}$. 

    By a standard induction on the iteration, we can show that $\sum_{j=1}^{i+1} x_j\geq OPT - (1-\frac{1}{k})^{i+1} OPT$. Plugging in $i+1 = \alpha k$, we get $\sum_{j=1}^{i+1} x_j\geq OPT - (1-\frac{1}{k})^{\alpha k} OPT \geq  (1-e^{-\alpha}) OPT$.
\end{proof}

We prove the following result using a similar technique as Lemma~\ref{lem: grdimm-strm}.
\begin{lemma}
\label{lem: grdimm-res-grdy}
The approximation ratio of  \greedimm\ with truncated greedy in local machines and streaming aggregation is $\frac{(1-e^{-\alpha}) (1/2-\delta)}{(1-e^{-\alpha}) +(1/2-\delta)} -\varepsilon$ in expection. Here, $0 \leq \alpha \leq 1$ is the fraction of $k$ local seeds communicated to the global machine.
\end{lemma}

\noindent{\bf Extension to other \emph{RIS}-based \infmax{} methods:}
Our \greedimm{} approach also extends to OPIM \cite{tang2018online}, which is another RIS-based approach to \infmax.
Unlike IMM, OPIM delivers an instance-wise approximation guarantee in each computation round and is suitable for online processing of \infmax{}. The instance-specific quality guarantees are achieved through partitioning the samples generated in each round (using the same \texttt{Sample(.)} subroutine as IMM) into two halves $R_1$ and $R_2$. The first is used to select an intermediate solution for that round (same mechanism as the \texttt{SelectSeeds(.)} step in IMM). The utility of this solution on $R_2$ is used as a validation score that is utilized by OPIM's version of the \texttt{CheckGoodness} subroutine to come up with the approximation guarantee for that round. 

Algorithmically, OPIM bears many similarities to IMM since it proceeds in rounds consisting of a sampling and seed selection phase and, as such, is also a suitable candidate for \greedimm. We include  experiments integrating OPIM with \greedimm{} (\S\ref{sec:opimeval}).

\subsection{Parallelization and implementation}
\label{sec:parallel}

\begin{figure}[tb]

    \centering
    
    \includegraphics[scale=0.5]{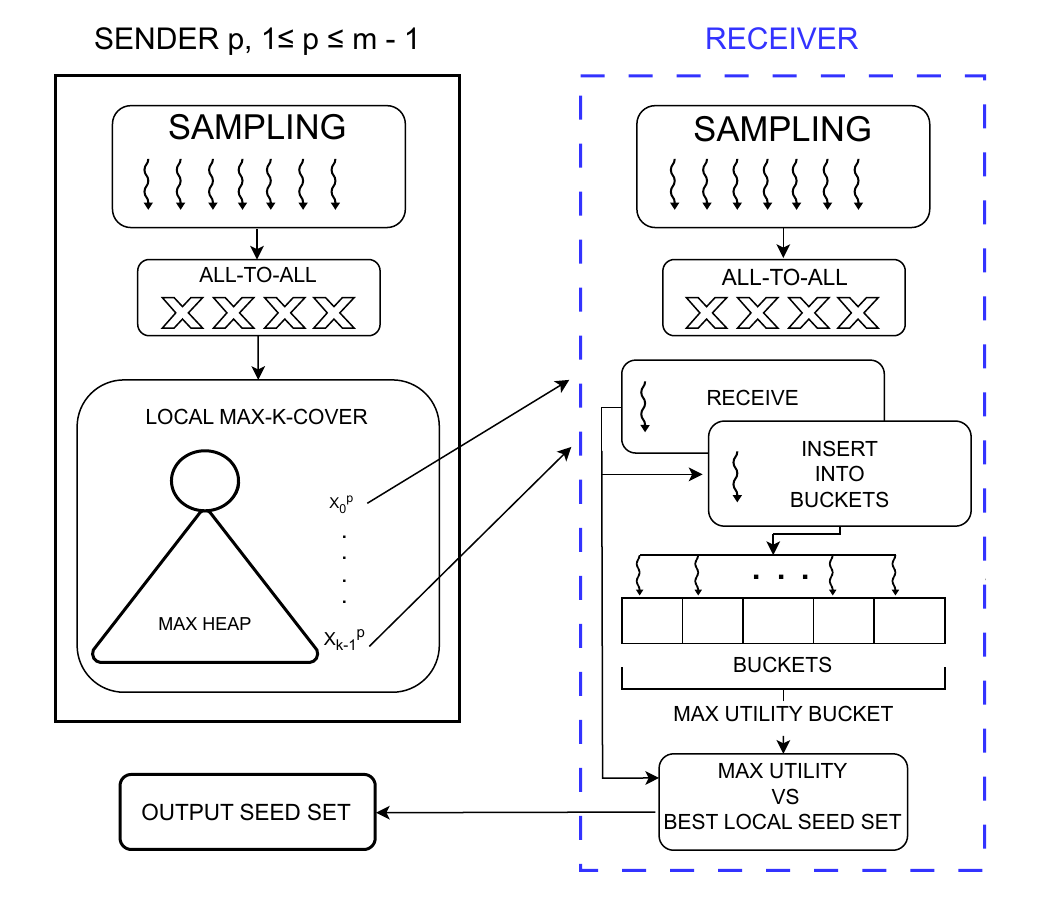}
    
    \caption{\footnotesize Schematic illustration of our parallel \greedimm{} approach inside one round.
    }
    \label{fig:greedimm}

\end{figure}

Next, we present the parallelization details for the overall \greedimm{} distributed streaming workflow. A schematic illustration of our \greedimm{} workflow is shown in Figure~\ref{fig:greedimm}.

\begin{compactenum}[S1)]
\item{\underline{Distributed sampling:}}

The input graph $G=(V,E)$ is loaded on all machines. Using $G$, all machines generate $\approx \theta/m$ samples each. This step supports multithreaded parallelism within each machine.

\item{\underline{All to all:}}

The list of vertex ids $[0,n-1]$ is partitioned uniformly at random so that each rank $p$ is assigned $\approx n/m$ ids, represented by $V_p$. Subsequently, using an MPI \texttt{all-to-allv} communication, we transport all RRR set ids that contain any given vertex $v\in V_p$ to rank $p$. At the end of this step, each machine stores the RRR set ids corresponding to all vertex ids in its assigned partition ($V_p$).
Recall the use of $\cS(u)$ for this covering set for each vertex id $u\in V_p$.
Henceforth, we distinguish between a ``sender'' process and a ``receiver'' process.
In our streaming implementation, there are $m-1$ senders (ranks $[1,m-1]$) and one receiver process (rank 0).

\item{\underline{Sender process:}}

A sender $p$ uses the collection of its $\approx n/m$ covering sets to detect up to $k$ local seeds. Let $[x_0^p,x_1^p,\ldots x_{k-1}^p]$ denote these seeds. 

We use the lazy greedy \maxkcover{} algorithm (Algorithm~\ref{lz-max-k-cover}) that internally maintains a max heap to identify the next best seed, starting with $x_0^p$.
To support streaming, the sender process sends each seed $x_i^p$ along with its covering subset to the receiver process as and when it is identified. This is implemented using a nonblocking send to minimize waiting at the sender process.
The sender concludes when all of its $k$ seeds are generated. Upon completion, it alerts the receiver of its termination.

For the truncated version (\S\ref{sec:truncated}), when the truncated limit is reached no more sends happen from that sender. But the sender continues to generate all its $k$ seeds locally. The idea is to enable comparison at the end with the global solution.

\item{\underline{Receiver process:}}

After participating in the initial sampling, the role of the receiver switches to: a) aggregating all the seeds sent by all senders; and b) performing the global \maxkcover{} using the multi-threaded streaming algorithm in Algorithm~\ref{alg:stream-cover}. 
Let $t$ denote the number of threads at the receiver. 
Thread rank 0 is a dedicated \textit{communicating thread} to listen to the senders using a \emph{nonblocking receive}. 
The remaining  $(t-1)$ threads parallelize the insertion of seeds into the $B$ buckets.  We refer to these as \textit{bucketing threads}.
We assign $\lceil B/(t-1)\rceil$ threads per bucket.
When a new seed $x$ (along with its covering set $\cS(x)$) is received, the communicating thread places $\langle x,\cS(x)\rangle$ on a queue and atomically sets a flag corresponding to that index. 
For this purpose, the receiver maintains a shared array $A$ of maximum size $m\cdot k$.
Each bucketing thread monitors the \emph{flag} corresponding to the next index yet to be read. When the flag is set, that means there is a new element $x$ in that index, then all bucketing threads read $x$ and insert them into their respective buckets independently. This allows for a lock-free  (atomic) update. 
This also allows concurrent processing of the bucketing threads with the communicating thread.
Finally, when all senders have alerted the receiver with termination, the receiver compares the best solution across the buckets with the best local solutions at the senders, and outputs the final seed set. 

\end{compactenum}

\noindent{\bf Runtime complexity:}
As for the runtime complexity, step S1 (sampling) takes $\mathcal{O}(\theta/m)\cdot\ell_s$, where $\ell_s$ is the average length of a sample.
Given the same set of edge probabilities, this average is expected to be more for the IC model than for LT.
Step S2 is an all-to-all communication step and can take $\mathcal{O}(\tau m + \mu\frac{n}{m}\theta)$ in the worst-case, where $\tau$ and $\mu$ are the network latency and reciprocal of bandwidth respectively. 
Note that the size of each covering set to be received is upper bounded by $\theta$ and in practice is likely to be less  (depends on its $n/m$ vertex subset).
Also, even though the receiver process sends out the samples it generated, it will not receive any data (from the all-to-all) as all the local \maxkcover{} happen at the senders.  
The  remaining part of the sender  process (S3) is consumed in generating local seeds from its local ($n/m$) covering subsets. Using the greedy algorithm, this takes $\mathcal{O}(\frac{k\theta n}{m})$.

As for the receiver process (S4), the communication time is dominated by the nonblocking receive to collect all the streamed seeds, and the computation time by the time to insert into the local $B$ buckets. However, considering masking and the parallel processing of buckets, we expect communication (or wait time for data to arrive) to dominate the total time---which is desirable to ensure the receiver is available for all senders. The $u/l$ ratio of the number of buckets in the streaming algorithm (\S\ref{sec:distIMM-streaming}) can be shown to be $k$. This is because the optimal cover could be at most $k$ times of the cover of a set with the maximum marginal gain. Using this,  the runtime of the streaming computation in the receiver is $\mathcal{O}(mk\theta \cdot B)=\mathcal{O}\left(mk\theta \cdot (\log_{(1+\delta)}{k})\right)$. \\

\noindent{\bf Memory complexity:}
Here we only discuss the memory complexity of the streaming algorithm. Each of the bucket needs to store the total cover of its partial solution. Since total cover could be at most $\theta$, we require $\mathcal{O}(\theta)$ memory per bucket. Each bucket also needs to store the partial solution of size $\mathcal{O}(k)$. Putting these together, the total memory complexity of the streaming algorithm in the receiver is $\mathcal{O}\left((\theta+k) \cdot \log_{(1+\delta)}{k}\right)$. 
This is more space efficient than any offline algorithm that would require storing the complete incoming data, i.e., $\mathcal{O}(mk\theta)$ elements. \\

\noindent{\bf Software availability:}
\greedimm{} is implemented using C/C++, MPI and OpenMP, and is open-sourced at the \href{https://github.com/ReetBarik/GreediRIS}{GreediRIS Github repository} \cite{GreeDIMMSoftware}.

\section{Experimental Evaluation}
\label{sec:results}

\subsection{Experimental setup}
\label{sec:expsetup}

\noindent{\bf Platform:}
All experiments were conducted on the NERSC Perlmutter supercomputer, a HPE Cray EX system. For our experiments, we used the CPU nodes (machines), each of which consists of two AMD EPYC 7763 (Milan) CPUs (with 64 cores per CPU), 256MB L3 cache, 512 GB of DDR4 memory, and a HPE Slingshot 11 NIC. 
Experiments were performed using up to $m=512$ nodes (with a total of 32K cores). 
All runtimes are measured in terms of wall clock time and reported in seconds, and all distributed runs were executed by binding one MPI rank per node. \\

\noindent{\bf Input data:}
We use nine real world networks from SNAP \cite{snapnets} and KONECT \cite{kunegis2013konect} summarized in Table~\ref{tab:Inputs}. 
The inputs cover a wide range in sizes, and application domains such as social networks, citation networks and web documents. 
Since edge probabilities are not available for these public networks, consistent with practice \cite{minutoli2019fast,minutoli2020curipples,DBLP:journals/tkde/LiFWT18}, we generated edge probabilities from a uniform random distribution between $[0,0.1]$.
Note that the weighted cascade (WC) model (i.e., edge weight from node $u$ to $v$ is given by $1 / InDegree(v)$)---as has been used in \cite{tang2022distributed}---is 
\emph{not} an option as it has been shown \cite{arora2017debunking} to \emph{not} translate to more generic diffusion models. \\

\begin{table}[tbh]
\footnotesize
\caption{\small Key details for our test inputs SNAP \cite{snapnets} and KONECT \cite{kunegis2013konect}). Avg. is the average out-degree and Max. is the maximum out-degree. 
}

\centering
\begin{tabular}{|l|r|r|r|r|}
\hline
\textbf{Input} & \textbf{\#Vertices} & \textbf{\#Edges} & \textbf{Avg.} & Max. \\ 
\hline
Github & 37,700 & 285,000 & 7.60 & 9,446 \\ 
HepPh & 34,546 & 421,578 & 24.41 & 846 \\ 
DBLP & 317,080 & 1,049,866 & 6.62 & 343 \\ 
Pokec & 1,632,803 & 30,622,564 & 37.51 & 20,518 \\ 
LiveJournal & 4,847,571 & 68,993,773 & 28.26 & 22,887 \\ 
Orkut & 3,072,441 & 117,184,899 & 76.28 & 33,313 \\ 
Orkut-group & 8,730,857 & 327,037,487 & 56.81 & 318,240 \\ 
Wikipedia & 13,593,032 & 437,217,424 & 22.56 & 5,576,228 \\  
Friendster & 65,608,366 & 1,806,067,135 & 27.528 & 3615 \\ 
\hline 
\end{tabular}
\label{tab:Inputs}
\end{table}

\noindent{\bf Evaluation methodology:}
Our experiments include both IC and LT diffusion models. 
For related work comparison, we compared our new implementations against two state-of-the-art distributed IMM methods, namely Ripples \cite{minutoli2019fast}  and DiIMM \cite{tang2022distributed}.
\textbf{Ripples} is available as an open source software \cite{ripplessoftware} and we compared with the C/C++ and OpenMP + MPI distributed implementation. 
The \textbf{DiIMM} software however was \textit{not} available as of this writing\footnote{We were not able to find its public repository and the authors did not respond despite multiple requests.}.
Therefore, to enable a comparison, we implemented our own version of their method into the Ripples open source package \cite{tang2022distributed}.
Additionally, we also added the functionality of using OPIM \cite{tang2018online} (instead of IMM) as the underlying \emph{RIS}-based \infmax{} strategy into \greedimm.

For our proposed approach, \greedimm, we tested two variants of our distributed algorithm:
 
\begin{itemize}[leftmargin=*,itemsep=-0.05ex]
    \item {\bf \greedimm}: uses the distributed streaming algorithm described in \S\ref{sec:distIMM-streaming} and its parallelization in \S\ref{sec:parallel};
    \item {\bf \greedimmtrunc}: uses the truncated extension of \greedimm, described in \S\ref{sec:truncated}, with $\alpha$  fraction of the  seeds communicated.
\end{itemize}

All runs not using OPIM were carried out for $k=100$ and  precision parameter $\varepsilon=0.13$.
For streaming,  we chose $\delta=0.077$---as this configuration set the number of buckets approximately equal to the number of available threads (63) at the global receiver. 
\greedimmtrunc{} was run for values of $\alpha\in(0,1]$. For the experiments using OPIM, we set $k = 1000$, $\varepsilon = 0.01$, and adjusted $\delta=0.0562$ to maintain the number of buckets at 63.

All implementations were compiled using (GCC 11.2.0; optimization flags \texttt{-O3} and \texttt{-mtune=native}), and MPI library Cray-mpich 8.1.24. 
For quality, we use the average number of node (vertex) activations over 5 simulations of the diffusion models (IC or LT) from the seed sets obtained by Ripples as the baseline, with the same for other implementations presented as a percentage change.

\subsection{Comparative evaluation}
\label{sec:comparativestudy}

Table~\ref{tab:perfVImmTime} shows comparative results in terms of runtime performance. \greedimmtrunc{} was the fastest for nearly all inputs, with \greedimm{} coming a close second. 
For all inputs tested, both \greedimm{} implementations significantly outperformed Ripples and DiIMM. 
For instance, for LT, the  speedups of \greedimmtrunc{} over Ripples ranged from $1.32\times$ (for friendster) to $357.78\times$ (for Github), with a geometric mean of $28.99\times$ across all inputs.
For IC, the corresponding speedups ranged from 
$1.38\times$ (for friendster) to $526.13\times$ (for Github), with a geometric mean of $36.35\times$ across all inputs.
The results also show that \greedimm{} benefits in runtime savings from truncation. 
Overall, these runtime results uniformly show the effectiveness of our \randgreedi-based distributed streaming as well as truncation for the seed selection step.
The variations of  speedups with inputs and models used is to be expected due to the effects of graph topology and stochasticity of the process.
Relative to IC, the performance benefits from \greedimm{} under LT is more because it has been known to generate shallower BFS traversals (i.e., shorter RRR set sizes).

\begin{table}[tbh]
\footnotesize
\caption{\footnotesize Performance of \greedimm{} 
 implementations,  Ripples \cite{minutoli2019fast} and our implementation of DiIMM \cite{tang2022distributed}. 
All results reported were using $m=512$ nodes and 32K cores in distributed memory. 
\greedimm-trunc  runs were performed using $\alpha=0.125$.
Bold values indicate the best entries for that input. 
}

\centering
\begin{tabular}{|l|rr|rr|} \hline
{\bf Diffusion: LT} & \multicolumn{4}{c|}{Time (in Sec.)} \\ \hline
Input & \multicolumn{1}{l}{Ripples\cite{tang2015influence}}  & \multicolumn{1}{l|}{DiIMM\cite{tang2022distributed}} & \multicolumn{1}{l}{\greedimm} & \multicolumn{1}{l|}{\greedimm-trunc} \\ \hline
Github & 108.3 & 114.7 & 1.8 & \textbf{0.2}  \\			
HepPh & 273.1 & 270.1 & 1.8 & \textbf{0.5} \\
DBLP & 365.0 & 357.7 & 3.7 & \textbf{3} \\
Pokec & 435.1 & 434.9 & 11.6 & \textbf{9.5} \\
LiveJournal & 482.1 & 478.0 & 22.0 & \textbf{22.9} \\
Orkut & 463.4 & 458.1 & 18.7 & \textbf{18.2} \\
Orkut-group & 549.4 & 564.6 & 81.7 & \textbf{78.6}\\
wikipedia & 537.0 & 528.4 & \textbf{45.0} & 46.6 \\
friendster & 948.2 & 994.9 & 746.1 & \textbf{685.9} \\ \hline

{\bf Diffusion: IC} & \multicolumn{4}{c|}{Time (in Sec.)}\\ \hline

Github & 122.0 & 112.0 & 0.9 & \textbf{0.3}\\
HepPh & 175.0 & 185.5 & 1.5 & \textbf{0.6} \\
DBLP & 266.0 & 274.7 & 2.4 & \textbf{1.1} \\
Pokec & 96.0 & 101.4 & 77.4 & \textbf{30.3}\\
LiveJournal & 129.0 & 146.4 & 100.0 & \textbf{64.3} \\
Orkut & 74.5 & 88.2 & 32.5 & \textbf{6.3} \\
Orkut-group & 164.3 & 181.1 & 119.1 & \textbf{60.6} \\
wikipedia & 360.3 & 287.5 & 256.0 & \textbf{222} \\
friendster & 278.0 & 319.0 & 361.0 & \textbf{211.2} \\ \hline
\end{tabular}

\label{tab:perfVImmTime}
\end{table}

We also compared the quality of seeds generated by \greedimm{} and  \greedimmtrunc, against the quality of seeds generated by Ripples. On average across all data sets, we observed that while using $m=512$ nodes, the expected influence achieved by the seeds generated by  \greedimm{} and \greedimmtrunc{} implementations was 2.72\% away from the influence generated by Ripples. 
This is despite reduced worst-case approximations (as shown in the Lemmas of \S\ref{sec:distIMM-streaming}) relative to Ripples which is $(1-1/e-\varepsilon)$-approximate.
For instance, our experimental settings for $\varepsilon = 0.13$ and $\delta = 0.077$ yield a worst-case approximation ratio of $0.123$ in expectation for \greedimm{} (compared to a $0.5$ ratio for Ripples).
These results show that despite being weaker in approximation guarantee, the practical quality of \greedimm{} and \greedimmtrunc{} are comparable  to Ripples, while providing significant performance advantage.

\subsection{Performance evaluation for \greedimm} 
\label{sec:perf}

Next, we present a detailed parallel performance evaluation of our \greedimm{} implementations. We first present the strong scaling results for our main \greedimm{} implementation. Table~\ref{tab:strongscaling} shows the  strong scaling results. In the interest of space, we show results using the IC model.
\begin{table}[tbh]
\centering
\small
\vspace{-0.025in}
\caption{\footnotesize {\em Strong scaling} performance of \greedimm{} for different inputs, varying $m$ up to $512$ nodes for the IC model. All times are  in seconds. Blank entries mean those runs were not performed. 
}
\begin{tabular}{l|rrrrrrr|}
\cline{2-8}
& \multicolumn{7}{c|}{Number of nodes ($m$)} \\ \cline{2-8} 
& \multicolumn{1}{r|}{\textit{8}} & \multicolumn{1}{r|}{\textit{16}} & \multicolumn{1}{r|}{\textit{32}} & \multicolumn{1}{r|}{\textit{64}} & \multicolumn{1}{r|}{\textit{128}} & \multicolumn{1}{r|}{\textit{256}} & \textit{512} \\ \hline
\multicolumn{1}{|l|}{Pokec} & \multicolumn{1}{r|}{594} & \multicolumn{1}{r|}{289} & \multicolumn{1}{r|}{152} & \multicolumn{1}{r|}{89} & \multicolumn{1}{r|}{62} & \multicolumn{1}{r|}{56} & 77 \\ \hline
\multicolumn{1}{|l|}{LiveJournal} & \multicolumn{1}{r|}{1,656} & \multicolumn{1}{r|}{796} & \multicolumn{1}{r|}{556} & \multicolumn{1}{r|}{208} & \multicolumn{1}{r|}{126} & \multicolumn{1}{r|}{106} & 100 \\ \hline
\multicolumn{1}{|l|}{Orkut} & \multicolumn{1}{r|}{71} & \multicolumn{1}{r|}{36} & \multicolumn{1}{r|}{18} & \multicolumn{1}{r|}{13} & \multicolumn{1}{r|}{14} & \multicolumn{1}{r|}{17.0} & 32 \\ \hline
\multicolumn{1}{|l|}{Orkut-group} & \multicolumn{1}{r|}{1,693} & \multicolumn{1}{r|}{837} & \multicolumn{1}{r|}{394} & \multicolumn{1}{r|}{208} & \multicolumn{1}{r|}{123} & \multicolumn{1}{r|}{108} & 119 \\ \hline
\multicolumn{1}{|l|}{wikipedia} & \multicolumn{1}{r|}{--} & \multicolumn{1}{r|}{--} & \multicolumn{1}{r|}{1,931} & \multicolumn{1}{r|}{984} & \multicolumn{1}{r|}{572} & \multicolumn{1}{r|}{414} & 256 \\ \hline
\multicolumn{1}{|l|}{friendster} & \multicolumn{1}{r|}{--} & \multicolumn{1}{r|}{--} & \multicolumn{1}{r|}{1,564} & \multicolumn{1}{r|}{866} & \multicolumn{1}{r|}{522} & \multicolumn{1}{r|}{414} & 361 \\ \hline
\end{tabular}
\label{tab:strongscaling}
\end{table}

Results for smaller inputs (Github, HepPh, DBLP) that took less than 3 seconds, are omitted.
On the remaining inputs, in general we see  better scaling behavior as the input size increases---e.g., scaling on LiveJournal is near-linear until $m=128$, while the further runtime reductions are continued to be achieved for up to $m=256$ with Orkut-group, and up to $m=512$ with Wikipedia and friendster.

\begin{wrapfigure}{l}
{0.60\textwidth}
    \centering
    \includegraphics[scale=0.47]{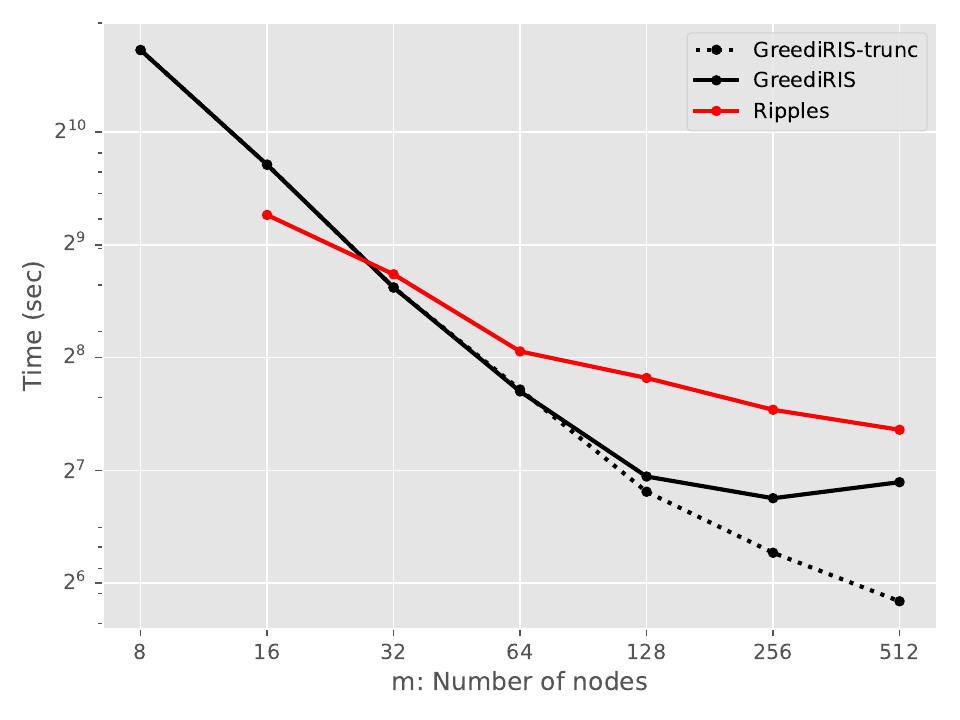}
    \vspace{-0.1cm}
    \caption{
    \footnotesize
    Scaling of the total execution time for our methods \greedimm{} and \greedimmtrunc{} on up to 512 Perlmutter nodes for input Orkut-group. Also shown is the scaling behavior of the state-of-the-art tool Ripples.}
    
    \label{fig:JugaadFigure}
\end{wrapfigure}

\begin{figure*}[hbt]
\centering
\begin{subfigure}[b]{0.49\textwidth}
    \centering
    \includegraphics[width=.9\textwidth]{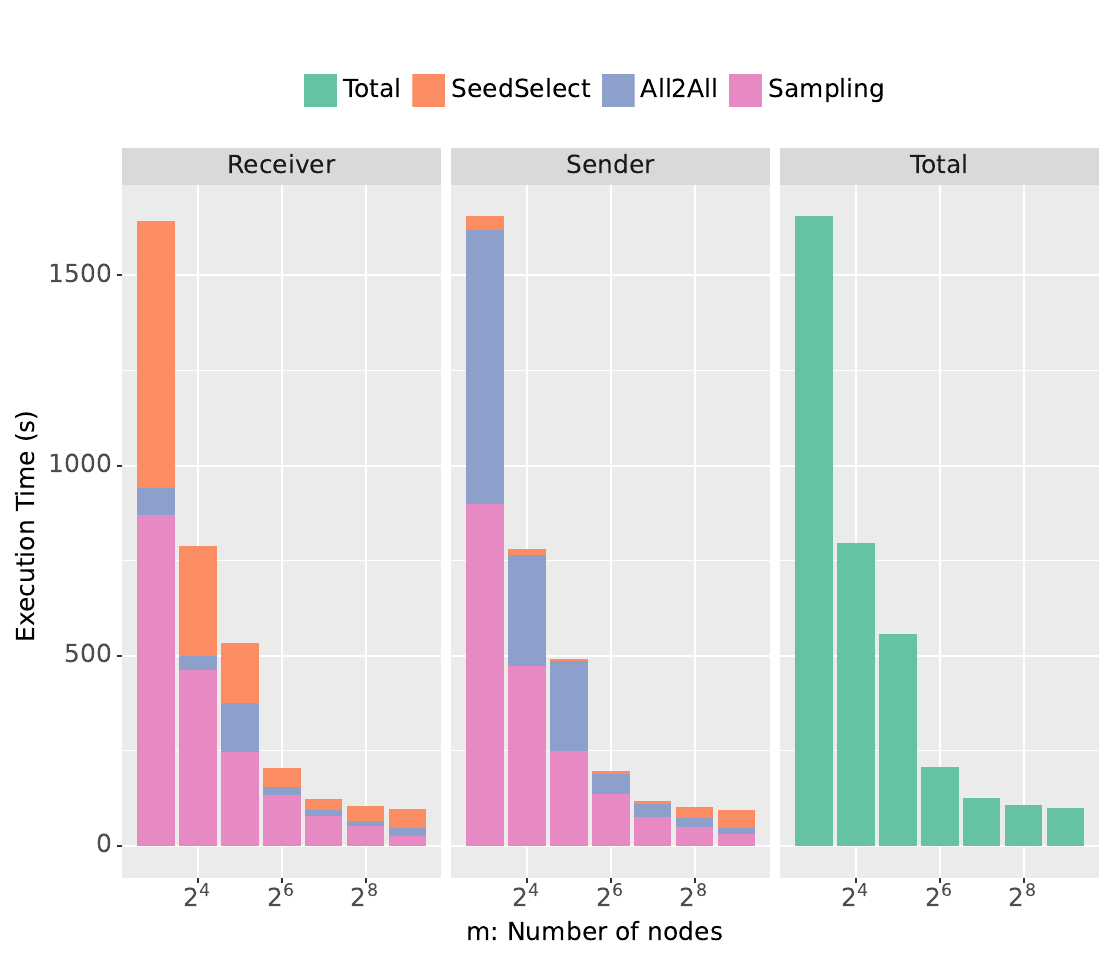}
    \caption{\footnotesize LiveJournal: Total time breakdown}
    \label{fig:lvjstrong:total}
\end{subfigure} 
\begin{subfigure}[b]{0.49\textwidth}
    \centering
    \includegraphics[width=.9\textwidth]{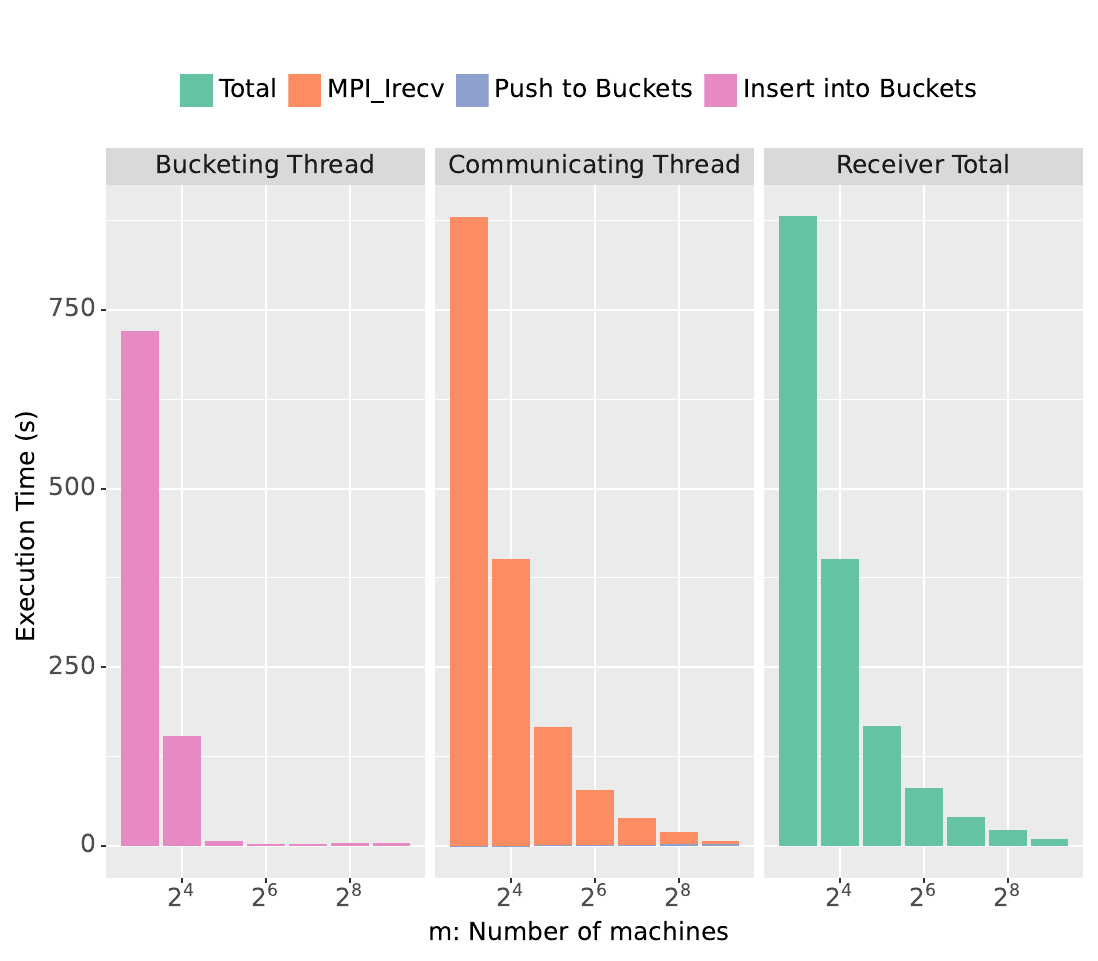}
    \caption{\footnotesize LiveJournal: Receiver time breakdown}
    \label{fig:lvjstrong:receiver}
\end{subfigure} 
    \caption{\footnotesize {\em Parallel runtime breakdown} for input LiveJournal (Diffusion model IC) for \greedimm: by the receiver, sender (longest running), and the total times. Note that in streaming, senders and the receiver run in parallel. The plot corresponding to the receiver shows the breakdown between its communicating thread and bucketing threads.
    Note that the majority of the SeedSelect time on the receiver is idle time, as senders participate in the all-to-all and then  perform their local seed selections. 
}
 \label{fig:totalbreakdown}
\end{figure*}

In general, we observed that \greedimm{} is able to push the scaling to larger number of nodes, beyond what Ripples achieves. This is illustrated in Figure~\ref{fig:JugaadFigure} where we see the scaling benefits of using \greedimm{} over Ripples for a representative input like Orkut-group. \greedimmtrunc{} helps push this boundary even further with more results shown at the end of this Section. 
Figure~\ref{fig:totalbreakdown} shows the detailed runtime breakdown for a representative input LiveJournal for the IC diffusion model. \\
For sender time, we used the time for the longest running sender.
As Figure~\ref{fig:lvjstrong:total} corresponding to the total breakdown shows, the total time is closer to the the maximum of the sender and receiver times than the sum, and these two times are nearly comparable --- suggesting the effectiveness of streaming. 
The sender times were evenly split between  sampling and all-to-all.
We see the sampling and all-to-all times scaling with $m$. 
The seed selection, however, starts to consume more time at the receiver for larger $m$ ($\geq 256$) settings. 

This is something the truncated version (\greedimmtrunc) is better equipped to address, as we will see next. 

We also examine the receiver process time closely for the same input in Figure~\ref{fig:lvjstrong:receiver}, since the global seed selection using streaming is carried out at the receiver.
Recall that at the receiver, thread rank 0 is the \textit{communicating thread}, as it monitors  the communication channel by doing a non-blocking receive, and when the next seed arrives, it pushes it to the local queue. 
All other $t-1$ threads (i.e., $63$) are \textit{bucketing threads}, handling insertions individually into a subset of $\lceil B/(t-1)\rceil$ buckets.
The results show that the communicating thread spends most of its time on the non-blocking receive, which implies high availability to the senders. 
The bucketing threads generally take significantly much less time, and their running times are subsumed within the communicating thread's time. Note that there are 63 threads to handle the buckets.

\noindent{\bf Evaluation of \greedimmtrunc{}: }

\begin{figure}[tbh]
    \centering
    \includegraphics[scale=0.60]{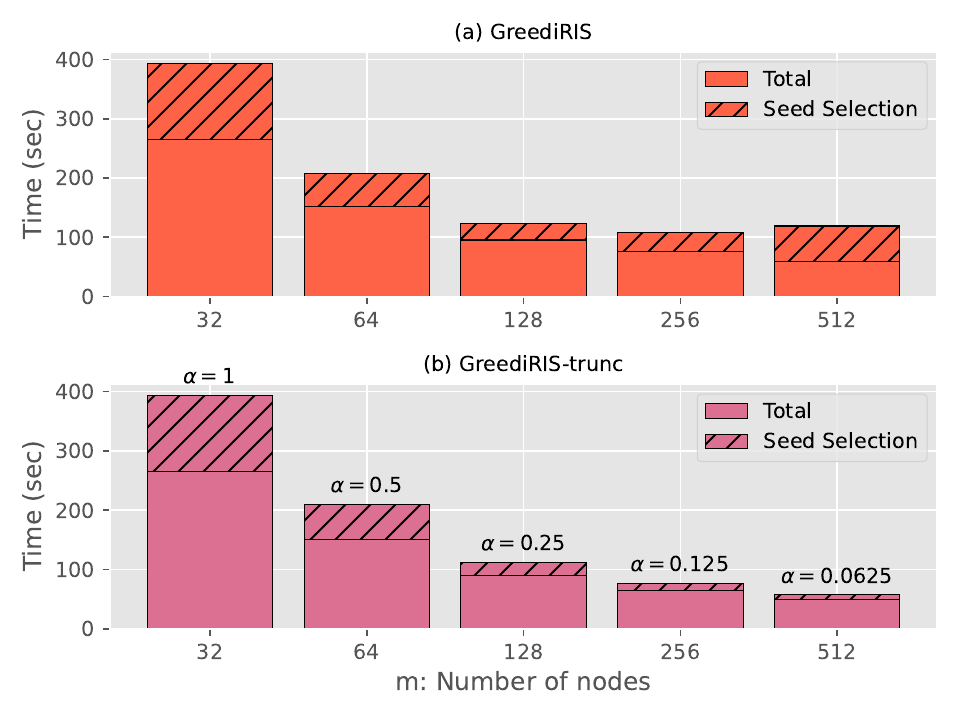}
    \vspace{-0.05cm}
    \caption{\footnotesize
    Strong scaling plot of \greedimm{} (top) and \greedimmtrunc{} (bottom) to up to 512 nodes. The seed selection step is shown as the shaded region representing its fraction of the total runtime.}
    
    \label{fig:trunc}
\end{figure}

In this section, we tested the framework's truncated variant, \greedimmtrunc{}  (\S\ref{sec:truncated}) by varying the parameter $\alpha$. Note that $\alpha$ is the fraction of seeds sent from each sender to the global receiver. Increasing the number of nodes leads to a proportional increase in the communication and computation performed by the global receiver.

The parameter $\alpha$ provides a way to cap this load on the receiver and extend scaling. This can be  seen in Figure~\ref{fig:trunc}(a), where the parallel runtime for \greedimm{} starts to plateau for \greedimm{} for $m\geq 256$, while for \greedimmtrunc{} it continues to decrease (mainly due to decrease in $\alpha$). 
While decreasing $\alpha$ results in a lower approximation gurantee (Lemma~\ref{lem: grdimm-res-grdy}) and consequently degrade quality, our experiments showed this to be negligible (less than $0.36\%$) in practice for any given $m$.

\subsection{Extension of \greedimm{} to OPIM}
\label{sec:opimeval}

In this section, we show OPIM results integrated to the \greedimm{} framework.

\begin{table}[tbh]
\caption{\footnotesize Evaluation of \greedimmtrunc{} using the OPIM RIS-strategy \cite{tang2018online}, and using 
parameters: \{$k = 1000$, $\varepsilon = 0.01$, target $\theta \approx 2^{20}$, $\delta=0.0562$\}. All runs were performed on the friendster input on $m=512$ nodes.
}

\centering
\begin{tabular}{|l|r|r|r|r|} 

\hline
Truncation factor $\alpha$: & \multicolumn{1}{|c|}{$1$} & \multicolumn{1}{|c|}{$0.5$} & \multicolumn{1}{|c|}{$0.25$} & \multicolumn{1}{|c|}{$0.125$}  \\ \hline
Seed select time (sec):  & 381.42 & 200.59 & 99.30	& 95.43  \\ \hline	

OPIM approx. guarantee: & 0.66 &	0.67 &	0.68 &	0.69 \\ \hline	
\end{tabular}

\label{tab:opimExp}
\end{table}

Consistent with the large scale experimental settings of \cite{tang2018online}, we set $k = 1000$, $\varepsilon = 0.01$, and terminate when the number of generated samples exceeds $2^{20}$. We set $\delta = 0.0562$ to ensure the number of buckets at the receiver is 63 (1 communicating thread vs. 63 bucketing threads). 
The number of nodes was set to $m=512$.
For evaluation, we used the \greedimmtrunc{} implementation that uses OPIM internally, and studied the seed-selection performance. Results are shown in Table~\ref{tab:opimExp}. As can be observed, we are able to achieve significant reduction in time with increasing $\alpha$, while maintaining the reported approximation guarantee (as reported by OPIM using the martingale based analysis in \cite{tang2018online}). 

\section{Conclusions and future work}
\label{sec:conclusions}

We presented \greedimm, a new scalable distributed streaming algorithm and its parallel implementations for \infmax. New ideas include a) leveraging the \randgreedi{} framework for distributed submodular optimization for \infmax, b) introducing streaming into distributed \maxkcover{} allowing efficient masking of communication overheads through overlapped computation; and c) truncation to further reduce communication burden. The experimental study demonstrated that  \greedimm{} significantly outperforms state-of-the-art distributed parallel implementations with comparable quality. The algorithms presented are generalizable to any other monotone submodular optimization problem. 

Future extensions and directions could include:
i) enabling streaming during the sampling phase to mask all-to-all communication costs; 
ii) further optimizations to increase problem size reach to solve larger problems on smaller systems;
iii) GPU acceleration for sampling; and
iv) extension to other monotone submodular optimization problems.
\section{Acknowledgement}
\label{sec:ack}
This work is in part supported NSF grants CCF 2316160 and CCF 1919122 to Washington State University. 
The work is also supported by the U.S. Department of Energy through the Exascale Computing Project (17-SC-20-SC) (ExaGraph) at the Pacific Northwest National Laboratory.

\bibliographystyle{elsarticle-num-names} 
\bibliography{references}

\end{document}